\documentclass[useAMS,usenatbib,usegraphicx]{mn2e}
\usepackage{times}

\title[]{Investigation of the single neutron exposure model for the s-process: the primary nature of the  neutron source}
\author[Kun Ma, Wenyuan Cui and Bo Zhang]{Kun Ma$^{1}$, Wenyuan
Cui$^{1,2}$ and Bo Zhang$^{1}$\thanks{Corresponding author.
Email-address: zhangbo@mail.hebtu.edu.cn (Bo Zhang)}\\
$^{1}$Department of Physics, Hebei Normal University, 113 Yuhua
Dong Road, Shijiazhuang 050016, China\\
$^{2}$National Astronomical observatories, Chinese Academy of
Sciences, Beijing 100012,  China}
\begin{document}

\date{Received...; in original form...}

\pagerange{\pageref{firstpage}--\pageref{lastpage}} \pubyear{2006}

\maketitle

\label{firstpage}

\begin{abstract}
The primary nature of the $^{13}$C neutron source is very
significant for the studies of the s-process nucleosynthesis. In
this paper we present an attempt to fit the element abundances
observed in 16 s-rich stars using parametric model of the single
neutron exposure. The calculated results indicate that almost all
s-elements were made in a single neutron exposure for 9 sample
stars. Although a large spread of neutron exposure is obtained,
the maximum value of the neutron exposure will reach about 7.0
mbarn$^{-1}$, which is close to the theoretical predictions by the
AGB model. The calculated result is a significant evidence for the
primary nature of the  neutron source. Combining the result
obtained in this work and the neutron exposure-initial mass
relations, a large spread of neutron exposure can be explained by
the different initial stellar mass and their time evolution. The
possibility that the rotationally induced mixing process can lead
to a spread of the neutron exposure in AGB stars is also existent.
\end{abstract}

\begin{keywords}
nucleosynthesis: abundances-stars: AGB-stars.
\end{keywords}

\section{Introduction}

The two neutron-capture processes, the slow (s-process) and the
rapid (the r-process), occur under different physical conditions,
and therefore are likely to arise in different astrophysical sites
 \citep{bu57}. The most likely site for the s-process is
the inter-shell region of a thermally pulsing AGB star, provided a
suitable neutron source is active. Our present understanding of
the behavior of s-process nucleosynthesis has been reviewed by
\citet{bu99}. There is a general consensus that the neutron source
is the reaction $^{13}$C($\alpha$, n)$^{16}$O. In order to
activate it, a partial mixing of protons (PMP) from the envelope
down into the C-rich layers is required. PMP activates the chain
of reactions ($^{12}$C(p,
$\gamma$)$^{13}$N($\beta$)$^{13}$C($\alpha$, n)$^{16}$O). The
physical cause of PMP could be diffusive convective overshooting
\citep{he97,go00}, rotationally induced mixing \citep{la99,he03}
and gravity waves \citep{de03}. The s-elements produced in the
deep interior by successive neutron captures are subsequently
brought to the surface by the third dredge-up \citep{ga98}. Using
the primary-like neutron source ($^{13}$C($\alpha$, n)$^{16}$O)
and starting with a very low initial metallicity, most iron seeds
are converted into $^{208}$Pb. So, when third dredge up episodes
mix the neutron capture products into the envelope, the star
appears s-enhanced and lead-rich. Therefore, if the standard PMP
scenario holds, all s-process-enriched AGB stars with
metallicities [Fe/H]$\leq$-1.3 are thus predicted to be Pb stars
([Pb/hs]$\geq$1, where ``hs" denotes the `heavy' s-process
elements such as Ba, La, Ce), independent of their mass and
metallicity \citep{go00}.

The first three such lead stars (HD187861, HD224959, HD196944)
have been reported by \citet{va01}. At the same time, \citet{ao01}
found that the slightly more metal-deficient stars LP 625-44 and
LP 706-7 are enriched in s-elements, but cannot be considered as
lead stars, in disagreement with the standard PMP predictions.

The large observation data spreads of [Pb/hs] are strong
indication to suspect a large intrinsic spread of integrated
neutron irradiations. A large spread of $^{13}$C pocket
efficiencies is proposed by \citet{st05} in order to explain the
spreads of [Pb/hs]. It should be stressed here that the primary
nature of the $^{13}$C neutron source are rather robust
\citep{go00}. In the framework of the PMP scenario, there is no
obvious degree of freedom that could be used to reduce the
$^{13}$C pocket efficiencies in low-metallicity AGB stars
\citep{va03}. The possibility that the rotationally induced mixing
process may lead to a spread of the neutron exposure in AGB stars
is proposed by \citet{he03}. \citet{si04} explored the effects of
rotationally induced mixing on the nucleosynthesis of s-elements
during the TP-AGB phase. The results show that rotation of the AGB
star quenches the s-process efficiency because of the
contamination of the $^{13}$C layer by $^{14}$N. They find that
although rotational mixing is an efficient mechanism to trigger
the third dredge-up, but this process fails to reproduce the
observed strong s-process overabundances. The dependence of the
induced mixing on the initial value of the rotational velocity is
also largely unknown. So the fundamental problems, such as the
formation and the consistency of the $^{13}$C pocket, the
development of the third dredge-up and the neutron exposure
signature in the interpulse, currently exist in the models of AGB
stars. In contrast to other studies, \citet{cu06} find that a
large spread of the $^{13}$C efficiency is not needed to explain
the observed spread of [Pb/hs], but this comes naturally from the
range of different initial stellar masses and their time evolution
\citep[see also][]{bo06}. The neutron exposure per circle deduced
for the s-rich stars lies between 0.45 and 0.88 mbarn$^{-1}$
\citep{zh06}. For the metallicites of the Pb-enhanced stars, based
on the primary nature of the $^{13}$C neutron source \citep{bu99},
the neutron exposure per interpulse will reach about 7.0
mbarn$^{-1}$ \citep{cu06} which are about 10 times of the results
obtained by \citet{zh06} for s-rich metal-poor stars.

Another possibility is that the s-process material has experienced
only a few neutron exposures in the convective He-burning shell.
This is consistent with a proposed mechanism for the s-process in
metal-poor AGB stars with [Fe/H]$<$-2.5 \citep{fu00}. These
authors proposed a scenario in which the convective shell
triggered by the thermal runaway develops inside the helium layer.
Once this occurs, $^{12}$C captures proton to synthesize $^{13}$C
and other neutron-source nuclei. The thermal runaway continues to
heat material in the thermal pulse so that neutrons produced by
the $^{22}$Ne($\alpha$, n)$^{25}$Mg reaction as well as the
$^{13}$C($\alpha$, n)$^{16}$O reaction may contribute. It is
possible that only one episode of proton mixing into He intershell
layer occurs in metal-poor stars \citep{fu00,ao01,iw02}. After the
first two pulses no more proton mixing occurs although the third
dredge-up events continue to repeat, so the abundances of the
s-rich metal-poor stars can be characterized by only one neutron
exposure. Detailed stellar evolution calculations are therefore
highly desired, in order to clarify which site is the most likely
to dominate the s-process in metal-poor AGB stars [interpulse
\citep{ga98}, or thermal pulse \citep{fu00}].

Obviously, the detailed study of s-rich stars are needed in order
to make progress in our understanding of the s-rich phenomenon,
investigate what its physical reasons might be and for
constraining what the possible physical conditions are. These
reasons motivated us to start a systematic study for s-rich stars.
In this paper we use the parametric approach of only one neutron
exposure to investigate the characteristics of the nucleosynthesis
pathway that produces the abundance ratios of s-rich objects.

\section{Parametric model of the single neutron exposure}

In order to investigate the efficiency of the s-process, the
elemental abundances of s-rich stars are particularly useful.
There have been many theoretical studies of s-process
nucleosynthesis in low-mass AGB stars. Unfortunately, however, the
precise mechanism for chemical mixing of protons from the
hydrogen-rich envelop into the $^{12}$C-rich layer to form
$^{13}$C-pocket ($^{12}$C(p,$\gamma$)$^{13}$N($e^+$,$\nu$)$^{13}$C
) is still unknown. In this paper we analyze the direct
observational constraints provided by the photospheric composition
of s-rich stars. Preliminary attempts to fitting their abundances
either using parameterized s-process distributions (e.g.
\citealt{ao01} ) give encouraging results. For this reason, we use
the parametric approach of only one neutron exposure, with many of
neutron-capture rates updated \citep{ba00}, to investigate what
physical conditions are possible to reproduce the observed
abundance pattern found in the s-rich stars
\citep{ao01,ao02,co03,jo02,lu03,ba05,iv05,jo06}.

We explored the origin of the neutron-capture elements in s-rich
stars by comparing the observed abundances with predicted s- and
r-process contribution. The \textit{i}-th element abundance in the
envelope of the star can be calculated as follows \citep{zh06}:
\begin{equation} N_{i}(Z)=C_{s}N_{i,\ s}+C_rN_{i,\ r}10^{[Fe/H]} ,
\end{equation}
where Z is the metallicity of the star, $N_{i,\ s}$ is the
abundance of the \textit{i}-th element produced by the s-process
in AGB star and $N_{i,\ r}$ the abundance of the \textit{i}-th
element produced by the r-process(per Si=$10^6$ at Z=Z$_\odot$),
$C_s$ and $C_r$ are the component coefficients that correspond to
relative contribution from the s-process and the r-process. It
should be noted that the s-process abundance in the envelope of
the stars could be expected to be lower than the abundance
produced by the s-process in AGB star because the material is
mixed with the envelopes of the primary (former AGB star) and
secondary stars.

For the single neutron exposure model, the overlap factor r, which
is the fraction of material that remains to experience subsequent
neutron exposures, is not a parameter because there is no material
that experience subsequent neutron exposures. So there are only
three parameters in the parametric model. They are the neutron
exposure $\Delta\tau$, the component coefficient of the s-process
C$_s$ and the component coefficient of the r-process C$_r$.

We can carry out s-process nucleosynthesis calculation of single
neutron exposure to fit the abundance profile observed in the
s-rich stars, in order to look for the minimum $\chi^2$ in the
three-parameter space formed by $\Delta\tau$, C$_s$ and C$_r$.
Using the method presented by \citet{ao01}, we chose Sr as the
representative for the first peak elements, Ba as the
representative for the second peak elements and Pb as the
representative for the third peak elements, so the uncertainties
of the parameters are determined by the error limits of the
representative elements. The adopted initial abundances of seed
nuclei lighter than the iron peak elements were taken to be the
solar-system abundances, scaled to the value of [Fe/H] of the
star. Because the neutron-capture-element component of the
interstellar gas that formed very mental-deficient stars is
expected to consist of mostly pure r-process elements, for the
other heavier nuclei we use the r-process abundances of the solar
system \citep{ar99}, normalized to the value of [Fe/H]. The
adopted abundances of r-process nuclei in equation (1) are taken
to be the solar-system r-process abundances \citep{ar99} for the
elements heavier than Ba, for the other lighter nuclei we use
solar-system r-process abundances multiplied by a factor of 0.4266
\citep{zh06}.

\section{Results and Discussion}

With the observed data in 16 sample stars, the model parameters
can be obtained from the parametric approach. The results of the
neutron exposures $\Delta\tau$, C$_s$, C$_r$ and s-process
fractional contributions for Ba and Eu, f$_{Ba,s}$ and f$_{Eu,s}$
are listed in the table 1. For 9 sample stars (the first 9 stars
listed in table 1), a single neutron exposure fits well the
observed data within the error limits of the representative
elements, whereas for others, although we can find the minimum
$\chi^2$ in the three-parameter space, a single neutron exposure
does not provide the parameters within the error limits for the
entire representative elements.

\begin{table*}
 \centering
 \begin{minipage}{140mm}
  \caption{Observed abundance ratios and the derived parameters
for s-rich stars}
  \begin{tabular}{@{}lccccclrccc@{}}
  \hline
 Star & [Fe/H] &[Ba/Fe]
&[Eu/Fe] &[Pb/Ba] &$\Delta\tau $  &$C_s$ &$C_r$
&$f_{Ba,s}$ &$f_{Eu,s}$ &$\chi^2$\\
&  & & & &(mbarn$^{-1}$)  & & & \%& \%&\\
 \hline
 CS 29526-110  &-2.38 &2.11 &1.73  &1.19
&4.10  &0.0050 &49.6 &89.5 & 14.5&0.726588\\
 CS 22898-027  &-2.25 &2.23 &1.88  &0.61
&0.88  &0.0035  &62.8  &90.6 & 23.3 &1.435540\\
 CS 31062-012  &-2.55 &1.98 &1.62  &0.42
&0.84  &0.0017  &34.7  &91.1 &22.9 &1.162923\\
 CS 31062-050  &-2.32 &2.30 &1.84  &0.60
&0.82  &0.0041 &49.3  &94.9 &34.0 &2.229041\\
HE 2148-1247  &-2.30  &2.36 &1.98  &0.76
&0.90  &0.0049  &64.5  &92.3 & 28.1 &2.130968\\
 LP 625-44  &-2.71  &2.74 &1.97  &-0.19
&0.70  &0.0058 &63.6  &96.7 &35.4 &3.006556\\
 LP 706-7  &-2.74  &2.01 &1.40  &0.27
 &0.84  &0.0019 &15.9  & 96.1 & 41.3 &0.602440\\
HE 0024-2523  &-2.71  &1.46 &1.10  &1.84
&7.12  &0.0014 &10.7  & 92.2 &19.7 &4.273823\\
HE 0338-3945  &-2.42  &2.41 &1.94  &0.69
&0.90  &0.0055  &55.5  &94.0 &33.8 &1.435311\\
CS 22948-27 &-2.47  &2.26 &1.88 &0.61
&0.76  &0.0026  &57.6 &92.7 &22.9 &0.990173\\
 CS 29497-34  &-2.90 &2.03 &1.80  &0.92
 &3.80 &0.0035  &56.9  &82.6 &8.5 &2.399216\\
 HD 196944  &-2.25  &1.10 &0.17  &0.80
&3.30  &0.0010 &0.5  & 99.2&69.0 &2.598222\\
CS 30301-015  &-2.64 &1.45 &0.2  &0.25
&3.76  &0.0007 &0.7  & 98.6 &58.3&3.644140\\
CS 29497-030  &-2.57 &2.32 &1.99  &1.33
&4.54 &0.0053 &86.3  & 85.2 &10.3 &2.459433\\
CS 30322-023  &-3.41 &0.54 &-0.51  &0.95
&3.92  &0.0002 &0.1  &99.3 & 72.2&1.576791\\
CS 22183-015  &-3.12 &2.09 &1.39  &1.08
&1.28 &0.0056 &21.5  &85.0& 17.0&1.291154\\
\hline
\end{tabular}
\end{minipage}
\end{table*}

Figure 1 shows our calculated best-fit results for the 9 sample
stars. In order to facilitate the comparisons of the theoretical
abundances with observations, the observed abundances of heavy
elements are marked by filled circles in these figures. We note
from the figure that, for most stars, the curves produced by the
model are consistent with the observed abundances within the error
limits. The agreement of the model results with the observations
provides a strong support to the validity of the parametric model
adopted in this work.

\begin{figure*}
 \centering
 \includegraphics[width=0.88\textwidth,height=0.5\textheight]{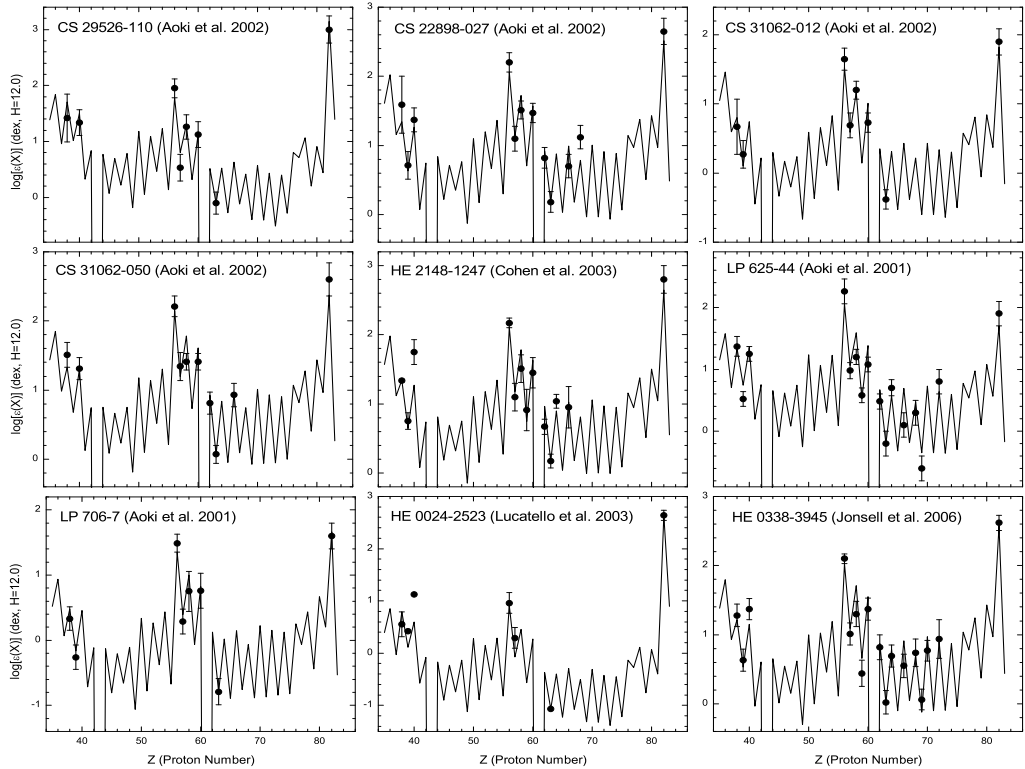}
 %\suppressfloats[t]
\caption{The Best fit to observational results of metal-deficient
stars. The black circles with appropriate error bars denote the
observed element abundances, the solid lines represent predictions
from s-process calculations considering r-process contribution.}
 %\label{appenfig}

\end{figure*}

The neutron exposure deduced for s-rich stars lies between 0.7 and
7.12 mbarn$^{-1}$. \citet{ao01} have reported a neutron exposure,
$\Delta\tau\sim 0.71 mbarn^{-1}$ for metal-poor star LP 625-44 and
$\Delta\tau\sim 0.8 mbarn^{-1}$ for LP 706-7, our calculated
results are close to their values. \citet{ga98} pointed out that
the neutron density is relatively low, reaching 10$^7$cm$^{-3}$ at
solar metallicity, corresponding to $\Delta\tau\sim0.2
mbarn^{-1}$. Since the $^{13}$C neutron source is of primary
nature, the typical neutron density in the nucleosynthesis zone
scales roughly as 1/Z$^{0.6}$, from Z$_\odot$ down to 1/50
Z$_\odot$. At lower metallicities, the effect of the primary
poisons prevails \citep{bu99}. For the metallicities of the sample
stars([Fe/H]$<$-2.0), based on the primary nature of the $^{13}$C
neutron source, the mean value of neutron density in the
$^{13}$C-pocket is about 35 times of the value for solar
metallicity \citep{ga99}, which indicates that the neutron
exposure per interpulse will reach about 7.0 mbarn$^{-1}$. It is
interesting to note that, although a large spread of neutron
exposure is shown in table 1, the maximum neutron exposure is
close to this value.

\begin{figure*}
 \centering
 \includegraphics[width=0.88\textwidth,height=0.5\textheight]{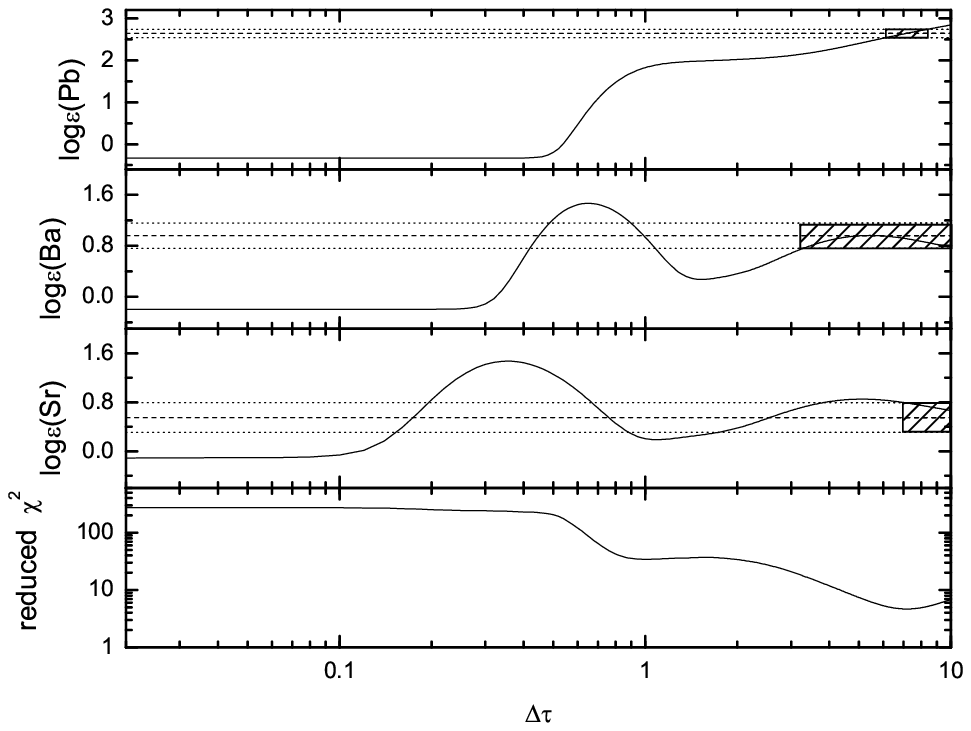}
 %\suppressfloats[t]
\caption{Calculated abundances log$\varepsilon$(Pb),
log$\varepsilon$(Ba),log$\varepsilon$(Sr) and reduced $\chi^2$, as
a function of the neutron exposure per pulse, $\Delta\tau$, in a
model with C$_r$=10.7 and C$_s$=0.0014. Solid curves refer to the
theoretical results, and dashed horizontal lines refer to the
observational results with errors expressed by dotted lines for HE
0024-2523. The shaded area illustrates the allowed region for the
theoretical model.}
 %\label{appenfig}

\end{figure*}
We discuss the uncertainty of the parameters using the method
presented by \citet{ao01}. Figure 2 shows the calculated
abundances log$\varepsilon$(Pb), log$\varepsilon$(Ba) and
log$\varepsilon$(Sr) as a function of the neutron exposure
$\Delta\tau$ in a model with C$_r$=10.7 and C$_s$=0.0014. These
are compared with the observed abundances of HE 0024-2523. There
is only one region in Figure 2,
$\Delta\tau$=7.12$^{+1.32}_{-0.10}$ mbarn$^{-1}$, in which all the
observed ratios of four representative elements can be accounted
for within the error limits. The bottom panel in Figure 2 displays
the reduced $\chi^2$ value calculated in our model with all
detected elemental abundances being taken into account and there
is a minimum, with $\chi^2$=4.27, at $\Delta\tau=7.12 mbarn^{-1}$.
For most sample stars shown in Figure 1, the uncertainties of the
neutron exposure are smaller than that of HE 0024-2523.

It is worth commenting on the behavior of log$\varepsilon$(Sr),
log$\varepsilon$(Ba) and log$\varepsilon$(Pb) as a function of the
neutron exposure $\Delta\tau$ seen in Figure 2. The nonlinear
trends displayed in the plot reveal the complex dependence on the
neutron exposure. The trends can be illustrated as follows.
Starting at low neutron exposure and moving toward higher neutron
exposure values, they show how the Sr peak elements are
preferentially produced at nearly $\Delta\tau\sim0.4 mbarn^{-1}$.
At larger neutron exposure (e.g., $\Delta\tau\sim0.7 mbarn^{-1}$),
the Ba-peak elements become dominant. Then a higher value of
$log\varepsilon(Pb)\sim2$ follows at $\Delta\tau=1.5 mbarn^{-1}$.
In this case, the s-process flow extends beyond the Sr-peak and
Ba-peak nuclei to cause an accumulation at $^{208}$Pb. At very
high neutron exposure values ($\Delta\tau$$>$2 mbarn$^{-1}$), the
role of Fe nuclei as seeds of the s-process is replaced partly by
lighter (intermediate atomic mass) nuclei. Neutron capture on them
can cross the iron peak, thus allowing the s-processing on heavy
isotopes to continue. It is clear from Figure 2 that there is
another maximum, with $log\varepsilon(Ba)=1.0$ at $\Delta\tau=5.5
mbarn^{-1}$. In this case, the Ba-peak elements become dominant
again. For extremely high neutron exposure values
($\Delta\tau$$>$6 mbarn$^{-1}$), the lighter seeds are also
converted into $^{208}$Pb. Clearly, log$\varepsilon$(Pb) is very
sensitive to the neutron exposure.

Our model is based on the observed abundances of the s-rich stars
and the nucleosynthesis calculations, so the uncertainties of
those observations and measurement of the neutron-capture cross
sections will be involved in the model calculations. For HE
0024-2523, abundances of four neutron capture elements (Sr, Ba, La
and Pb) and upper limits for an additional three elements
\citep[Y, Zr and Eu;][]{lu03} have been taken into account in our
model. The three upper limits will enlarge the uncertainties of
the parameters. We note from Table 1 that for four stars (CS
31062-50, HE 2148-1247, LP 625-44, HE 0024-2523), the reduced
$\chi^2$ are larger \ than 2. The probability that $\chi^2$ could
be this large as a result of random errors in the measurement of
the neutron-capture cross sections and abundances of the
neutron-capture elements is less than 2\%. We find that all these
uncertainties can not explain the larger errors of neutron-capture
elements, such as Zr in HE 2148-1247 and Y in LP 625-44. This
implies that our understanding of the true nature of s-process or
r-process is incomplete for at least some of these elements
\citep{tr04}.

It was possible to isolate the contributions corresponding to the
s- and r-process by our parametric model. In the Sun, the
elemental abundances of Ba and Eu consist of significantly
different combinations of s- and r-process isotope contributions,
with s:r ratios for Ba and Eu of 81:19 and 6:94, respectively
\citep{ar99}. The Ba and Eu abundances are most useful for
unraveling the sites and nuclear parameters associated with the s-
and r-process correspond to those in extremely metal-poor stars,
polluted by material with a few times of nucleosynthetic
processing. We explored the contributions of s- and r-process for
these two elements in the s-rich stars. In table 1 we display the
s-process fractions calculated from equation (1) for the 16 sample
stars. Clearly, for the first 9 stars listed in table 1, the s:r
ratios for Ba and Eu are larger than 90:10 and 15:85, which are
larger than the ratios in the solar system. The abundances of
r-elements, such as Eu, in s-rich stars are usually higher than
those in normal stars, so s-rich stars seemed to be also enriched
in r-elements, although in a lower degree than s-elements.

\section{Conclusions}

In this work, s- and r-process in s-rich stars were studied using
the parametric approach of single neutron exposure. Theoretical
predictions for abundances starting with Sr fit well the observed
data for 9 sample stars (the first nine stars listed in table 1),
providing an estimation for neutron exposure occurred in AGB
stars. The calculated results indicated that almost all s-elements
were made in the first neutron exposure for 9 sample stars. Once
this happens, after only one time dredge-up, the observed
abundance profile of the s-rich stars may be reproduced in a
single neutron exposure. This should be consistent with a proposed
mechanism for the s-process in metal-poor AGB stars with
[Fe/H]$<$-2.5 \citep{fu00}.

For the third dredge-up and the AGB model, several important
properties depend primarily on the core mass \citep{ib77,ka02}.
The analytical formula for AGB stars given by \citet{ib77} show
that the overlap factor decreases with increasing core mass.
\citet{ka02} and \citet{he00,he04} have also found that the third
dredge-up is more efficient for the AGB stars with larger core
masses. Since the core mass of AGB star at low metallicity is
larger than high metallicity obviously \citep{zi04}, the overlap
factor should be significantly smaller at low metallicity. This is
consist with the small overlap factor, r=0.1, deduced by
\citet{ao01} for metal-poor stars LP 625-44 and LP 706-7. In an
evolution model of AGB stars, a small r may be also realized if
the third dredge-up is deep enough for s-processed material to be
diluted by extensive admixture of unprocessed material. Clearly,
the behavior of very deep dredge-up (r=0) is also single neutron
exposure. Once this happens, no matter how many pulses may follow,
the observed abundance profile of the s-rich stars may be
reproduced in several similar single neutron exposures.

For the sample stars, the maximum value of the neutron exposure
obtained in this work reach about 7.0 mbarn$^{-1}$, which is close
to the theoretical predictions by the AGB model \citep{bu99} and
the maximum value of the case of radiative $^{13}$C-burning in the
low mass AGB stars \citep{cu06}. So the calculated result should
be a significant evidence for the primary nature of the neutron
source.

The large spread of neutron exposure given in table 1 could be
explained when varying the initial mass that affect dredge-up
events and nucleosynthetic processes along the AGB evolution
\citep{cu06,bo06}. Certainly, the possibility that the
rotationally induced mixing process may lead to a spread of the
neutron exposure in AGB stars is also existent. For intermediate
mass stars, the rapid rotation and significant amount of shear at
the bottom of the convective envelope in AGB stars are hard to
avoid. The rotations of the AGB star will quench the s-process
efficiency and reduce the neutron exposure. Through a distribution
of initial rotation rates, it may be lead to a natural spread of
neutron exposure in AGB stars. In table 1, one interesting result
is that, based on the values of the neutron exposure, the sample
stars should be divided into two groups: one group (include 8
sample stars) concentrate on the value of about 0.8 mbarn$^{-1}$
which correspond to the value of 2.5M$_\odot$ AGB star calculated
by \citet{cu06}, and the others (include 8 also sample stars) are
in the range of 1.3$-$7.12 mbarn$^{-1}$ which correspond to the
range of 1.5$-$2.0 M$_\odot$. Should the observed abundance
pattern of the group one belongs to be polluted by the rapid
rotators or the AGB stars with initial mass of about 2.5$-$3.0
M$_\odot$? Obviously, large uncertainties still remain in this
topic. More in-depth theoretical and observational studies of
s-rich stars will reveal the characteristics of the s-process at
low metallicity, such as their initial rotation rate and initial
mass dependence, and the history of enrichment of s- and
r-elements in the early Galaxy.

\section*{Acknowledgments}

We are grateful to the referee for the very valuable comments and
suggestions which improved this paper greatly. This work has been
supported by the National Natural Science Foundation of China
under grant no.10673002.

 \bsp

\label{lastpage}

\end{document}